\documentclass[reprint,aps,prmaterials,amsmath,amssymb,superscriptaddress,floatfix]{revtex4-2}

\usepackage{amstext}
\usepackage{graphicx}

\usepackage{color}
\usepackage{xspace}% correct spacing after formulas from macros
\usepackage{dcolumn}% Align table columns on decimal point
\usepackage{bm}% bold math
\usepackage[dvipsnames]{xcolor}% color
\usepackage{soul}% highlighting
\usepackage{gensymb}% \degree Celcius - use "\celsius" in text mode

\makeatother

\begin{document}
\title{Robust superconductivity and the suppression of charge-density wave in  $\text{Ca}_{3}(\text{Ir}_{1-x}\text{Rh}_{x})_{4}\text{Sn}_{13}$ single crystals at ambient pressure}

\author{Elizabeth H.~ Krenkel}
\affiliation{Ames National Laboratory, Ames, Iowa 50011, USA}
\affiliation{Department of Physics \& Astronomy, Iowa State University, Ames, Iowa
50011, USA}

\author{Makariy~A.~Tanatar}
\affiliation{Ames National Laboratory, Ames, Iowa 50011, USA}
\affiliation{Department of Physics \& Astronomy, Iowa State University, Ames, Iowa
50011, USA}

\author{Sunil~Ghimire}
\affiliation{Ames National Laboratory, Ames, Iowa 50011, USA}
\affiliation{Department of Physics \& Astronomy, Iowa State University, Ames, Iowa
50011, USA}

\author{Kamal~R.~Joshi}
\affiliation{Ames National Laboratory, Ames, Iowa 50011, USA}

\author{Shuzhang~Chen}
\affiliation{Condensed Matter Physics and Materials Science Department, Brookhaven
National Laboratory, Upton, New York 11973, USA}
\affiliation{Department of Physics and Astronomy, Stony Brook University, Stony
Brook, New York 11794-3800, USA}

\author{Cedomir~Petrovic}
\affiliation{Condensed Matter Physics and Materials Science Department, Brookhaven
National Laboratory, Upton, New York 11973, USA}
\affiliation{Department of Physics and Astronomy, Stony Brook University, Stony
Brook, New York 11794-3800, USA}

\author{Ruslan~Prozorov}
\email[Corresponding author:]{prozorov@ameslab.gov}
\affiliation{Ames National Laboratory, Ames, Iowa 50011, USA}
\affiliation{Department of Physics \& Astronomy, Iowa State University, Ames, Iowa
50011, USA}

\date{16 August 2023}

\begin{abstract}
Single crystals of Ca$_3$(Ir$_{1-x}$Rh$_x$)$_4$Sn$_{13}$ (3-4-13) were synthesized by flux growth and characterized by X-ray diffraction, EDX, magnetization, resistivity and radio frequency magnetic susceptibility tunnel diode resonator (TDR) techniques.
Compositional variation of the Rh/Ir ratio was used to study the coexistence and competition between the charge density wave (CDW) and superconductivity. The superconducting transition temperature varies from approximately 7 K in pure Ir ($x=0$)
to approximately 8.3 K in pure Rh ($x=1$). Temperature-dependent electrical resistivity reveals monotonic suppression of the CDW transition temperature, $T_{\text{CDW}}(x)$. The CDW starts in pure Ir, $x=0$, with $T_{\text{CDW}}\approx40$~K and extrapolates roughly linearly to zero at $x_c=0.58$ under the dome of superconductivity. Magnetization and transport measurements show a significant influence of CDW on the superconducting and normal state. Vortex pinning is substantially enhanced in the CDW region, and the resistivity is larger in this part of the phase diagram. The London penetration depth is attenuated exponentially upon cooling at low temperatures for all compositions, indicating a fully-gapped Fermi surface. We conclude that a novel $\text{Ca}_3(\text{Ir}_{1-x}\text{Rh}_x)_4\text{Sn}_{13}$ alloy with coexisting/competing CDW and superconductivity, is a good candidate to look for a composition-driven quantum critical point at ambient pressure. 
\end{abstract}
\maketitle

\section{Introduction}

The family of materials with general formula R$_{3}$T$_{4}$X$_{13}$ was discovered in the 1980 and is  frequently referred to as 3-4-13 Remeika compounds  \cite{Remeika1980}. Here, $R$ stands for alkali, rare earth, or actinide metals, $T$ stands for transition metal, and $X$ can be Sn, Ge, or In.  In this large family of materials (more than 1200, see Ref.\cite{Gumeniuk2018} for review), cage-like structures and strong electronic correlations provide possibilities to alter both the electronic and lattice properties. These materials are studied as potential thermoelectrics;
heavy fermion behavior is found in Ce$_3$Co$_4$Sn$_{13}$ \cite{Evan2006} and Pr$_3$Os$_4$Ge$_{13}$ \cite{Ogunbunmi2020} and itinerant ferromagnetism in Ce$_3$Os$_4$Ge$_{13}$ \cite{Prakash2016}. Superconductivity with transition temperatures below 4~K is found in a plethora of compounds; see Table 3 in the topical review article \cite{Gumeniuk2018}. Of all the Remeika series, superconductivity with a rather high $T_c$ is found in Yb$_3$Rh$_4$Sn$_{13}$ (7-8~K) \cite{Remeika1980} and in the compounds with coexisting charge density waves and superconductivity in (Ca,Sr)$_3$(Rh,Ir)$_4$Sn$_{13}$ series (6 to 8.5~K). 

In the case of these CDW superconductors, the highest transition temperatures have been found in the vicinity of a quantum critical point accessed by pressure \cite{Goh2015} and/or by alloying Sr$_3$Rh$_4$Sn$_{13}$ ($T_{CDW}$~135~K) with Ca$_3$Rh$_4$Sn$_{13}$ in $(\text{Ca}_{x}\text{Sr}_{1-x})_{3}\text{Rh}_{4}\text{Sn}_{13}$ series at around $x=0.9$  \cite{Klintberg2012,Cheung2018}. CDW order is suppressed in Ca$_3$Rh$_4$Sn$_{13}$ and superconducting $T_c$ reaches 8.3~K \cite{Biswas2014}.  
 The structural nature of QCP was confirmed by x-ray diffraction \cite{Veiga2020}.
The suppression of CDW reveals non-Fermi-liquid behavior of electrical resistivity,  evidencing strong fluctuations in the vicinity of the CDW boundary.  

Alloying Ca$_3$Ir$_4$Sn$_{13}$ ($T_{\text{CDW}}\sim$40~K) with 
$\text{Ca}_3\text{Rh}_4\text{Sn}_{13}$ provides a promising alternative means that could tune the system to CDW quantum critical point. Here, we report the synthesis and characterization of single crystals of $\text{Ca}_3(\text{Ir}_{1-x}\text{Rh}_x)_4\text{Sn}_{13}$. We find that the superconducting transition temperature remains constant in a range of compositions outside the CDW domain, rather than having a dome shape peaking at the QCP. 

\begin{figure}[tbh]
\includegraphics[width=7cm]{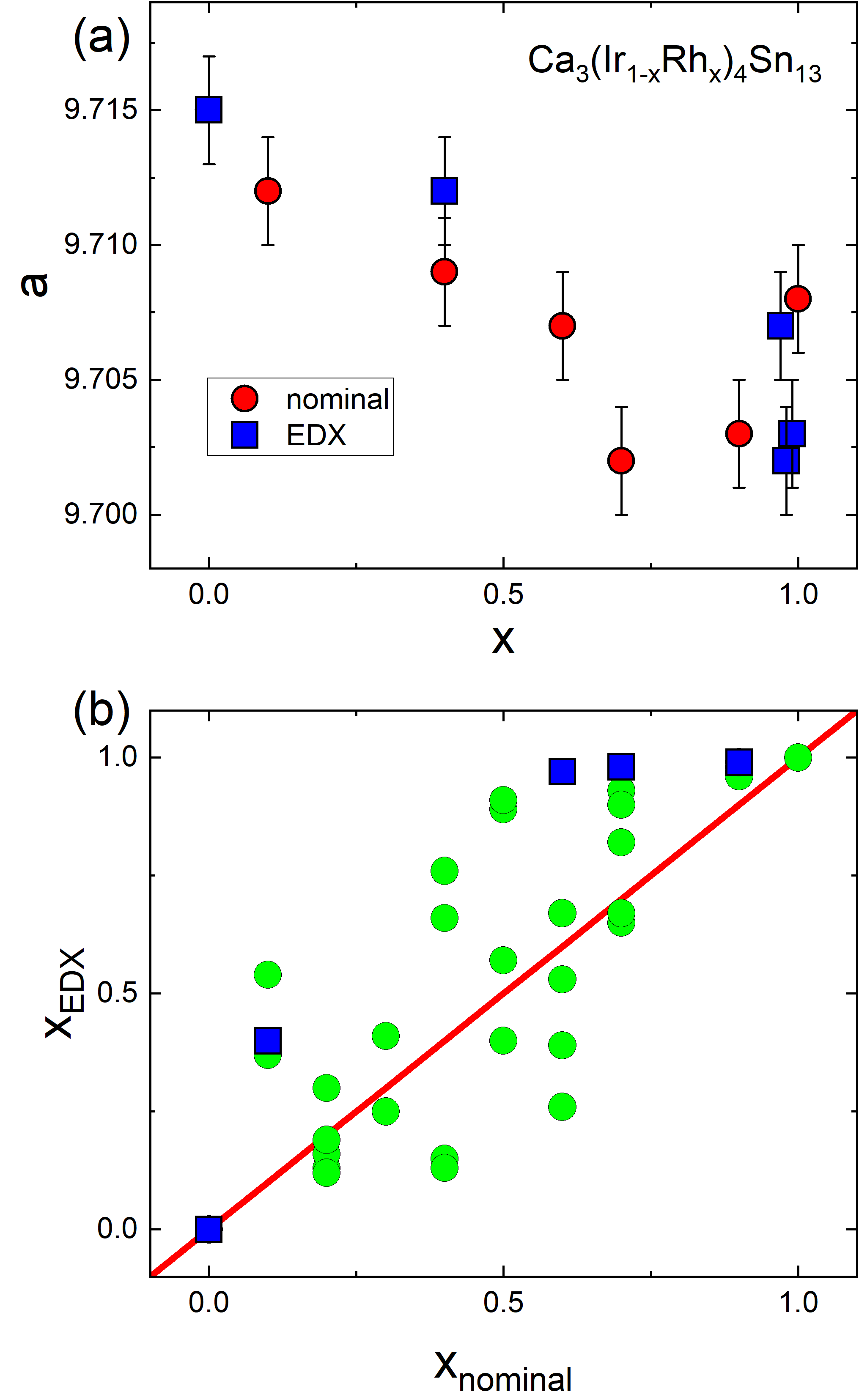} \caption{(a) Cubic lattice parameter $a$ as a function of nominal and EDX-determined
composition. (b) Composition determined from EDX, $x_{EDX}$ as function
of the nominal composition, $x_{nom}$. Violet squares show samples
used in panel (a). }
\label{fig1:x-ray} 
\end{figure}

\begin{figure*}[tbh]
\includegraphics[width=14cm]{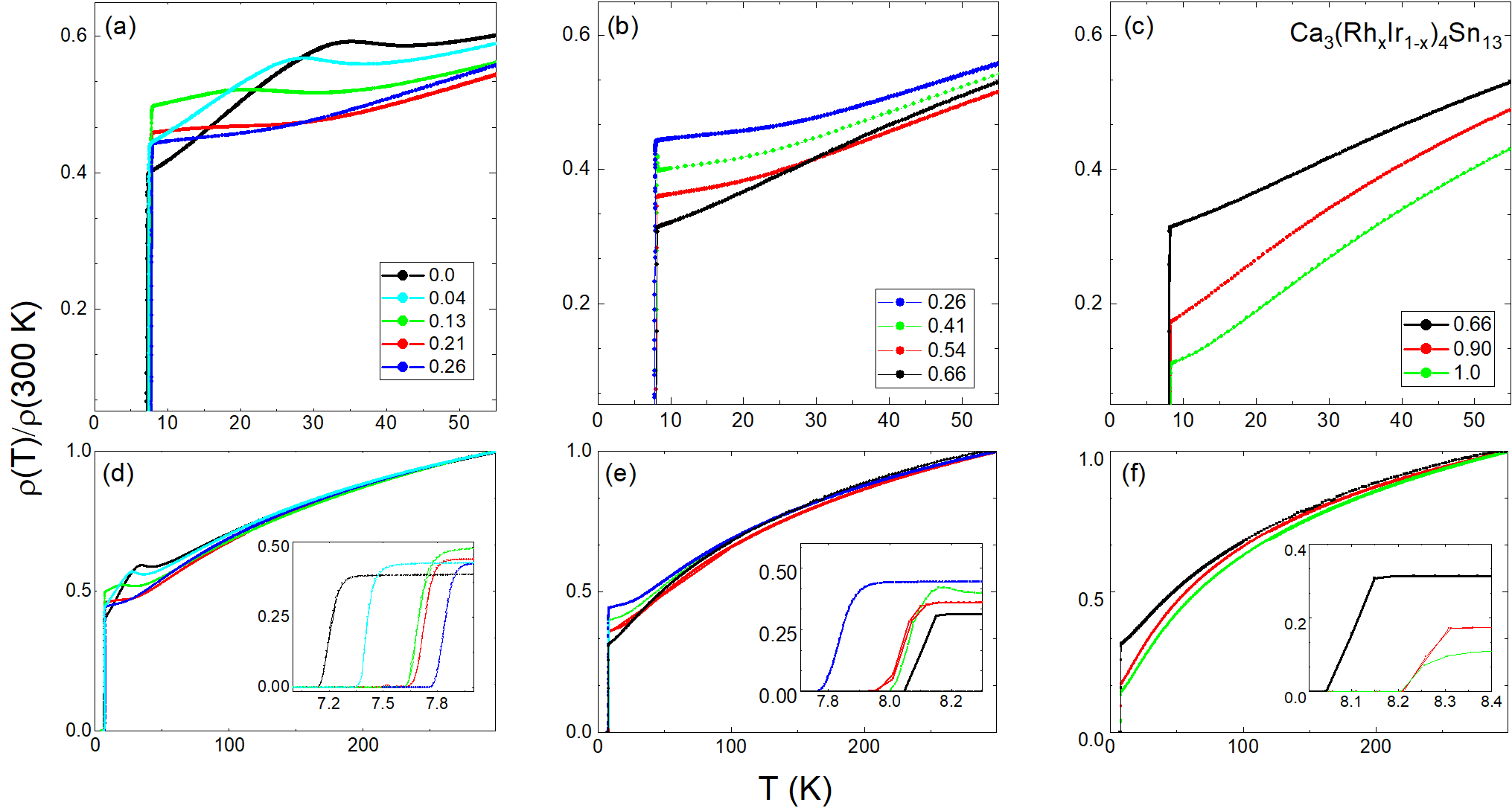} \caption{Temperature-dependent resistivity in $\text{Ca}_{3}(\text{Ir}_{1-x}\text{Rh}_{x})_{4}\text{Sn}_{13}$
single crystals normalized by its value at room temperature. (a,d)
Ir-rich compositions, (b,e) intermediate compositions, and (c,f) Rh-rich
compositions. Panels (a), (b) and (c) focus on the charge-density
wave (CDW); panels (d), (e) and (f) show the extended temperature
range up to room temperature. The insets focus on the superconducting
transition.}
\label{fig2:res}
\end{figure*}

\section{Experimental results}

\subsection{Crystal growth and chemical composition}

Single crystals of Ca$_3$(Ir$_{1-x}$Rh$_x$)$_4$Sn$_{13}$ were grown using a
high temperature self-flux method \cite{Remeika1980,cairsn,cairsn1}. Pieces of Ca, M and Sn (M = Ir, Rh) lumps were mixed in 3 -4 - 93 ratio, where M elements were pre-arc melted in desired stoichiomety for homogenization. The mixture
was placed in an alumina crucible, flushed with high purity argon
gas, and vacuum sealed in a quartz tube. The ampoule
was heated to 1100 $^{\circ}$C in a furnace, and kept over 6 h before
cooling to 800 $^{\circ}$C over 30 hours and over 140 hours down to 490 $^{\circ}$C. Shiny crystals with a typical size of few millimeters were obtained
after crystal decanting in a centrifuge. Extra residual Sn
flux was cleaned by polishing before any measurements. X-ray diffraction
(XRD) data were taken with Cu K$_{\alpha}$ ($\lambda=0.15418$ nm)
radiation of Rigaku Miniflex powder diffractometer. Figure \ref{fig1:x-ray}(a)
shows the lattice parameter $a$ as a function of the fractional composition
of Rh.

The actual composition was measured using JEOL SEM (scanning electron
microscope) equipped with EDX (Energy Dispersive X-ray spectroscopy)
detector. Each sample was characterized individually, as 
significant composition variation was found within batches. In each sample, several different
spots were examined. Although some crystals showed extremely high variation within the crystal, we selected crystals with a relative variation of less than 7\%. Therefore, for all resistivity, TDR, and magnetization measurements
reported in this paper, each crystal was measured with EDX.
This is necessary when studying $\text{Ca}_{3}(\text{Ir}_{1-x}\text{Rh}_{x})_{4}\text{Sn}_{13}$, because the system tends to grow a Rh-rich phase alongside other crystals, which makes the nominal composition extremely inaccurate.  Figure~\ref{fig1:x-ray}(a) shows the cubic lattice parameter, $a$, as a function of nominal and EDX-determined compositions. The same set of EDX compositions is shown by blue squares in Fig.~\ref{fig1:x-ray}(b). 
Panel (b) also shows many more crystals. The solid line shows $x_{\text{EDX}}=x_{\text{nom}}$ to guide the eye.  We can see that there is some trend in the data that follows the expected composition, but the scatter is significant.  In this work, the composition of EDX was determined for each sample shown, and we find systematic variations in properties as a function of $x_{\text{EDX}}$.

\subsection{Electrical resistivity}

Electrical resistivity was measured on single crystals shaped into ``resistivity bars'' for four-probe measurement. The size of the as-grown crystals ranged from sub-millimeter to 5 mm. The crystals were cut with a wire saw and polished to a typical size $(1-2)\times0.2\times0.4~\text{mm}^{3}$. The contacts were formed by soldering 50 $\mu$m silver wires with tin-silver solder \cite{SUST,Krenkel2022}, with the typical contact resistance below 100 $\mu\Omega$.

The resistivity of the samples at room temperature
$T=300~$K ranged between 140 and 105 $\mu \Omega$cm \cite{Krenkel2022}. AC resistivity measurements were performed on a 9 T \textit{Quantum Design} physical property measurement system (PPMS). While cubic in the normal state, the Ir-rich compounds undergo a structural distortion associated with the charge density wave transition along $q=(0,1/2,1/2)$ ordering vector \cite{Klintberg2012} \cite{Mazzone2015}. However, the  superstructure doubles the unit cell and either leaves the lattice body centered cubic or leads to formation of three equivalent tetragonal domains in the twinned state both of which should be macroscopically isotropic. Our measurements in different orientations did not reveal a noticeable difference. Therefore, measurements presented in this work do not follow any particular orientation.

Figure.~\ref{fig2:res} shows the evolution of the temperature-dependent
resistivity in $\text{Ca}_{3}(\text{Ir}_{1-x}\text{Rh}_{x})_{4}\text{Sn}_{13}$ single crystals. To avoid crowding, %visual mess
the figure is split into three panels: (a,d) Ir-rich compositions, (b,e) intermediate compositions, and (c,f) Rh-rich compositions. Panels (a), (b) and (c) show lower temperatures where charge-density wave (CDW) is well resolved whereas panels (d), (e) and (f) show the full temperature range. The insets focus on the superconducting transitions. To avoid complications related to the uncertainty of the geometric factor, all data are shown normalized by the value at the room temperature.

\subsection{DC magnetization}

\begin{figure}[tb]
\includegraphics[width=7cm]{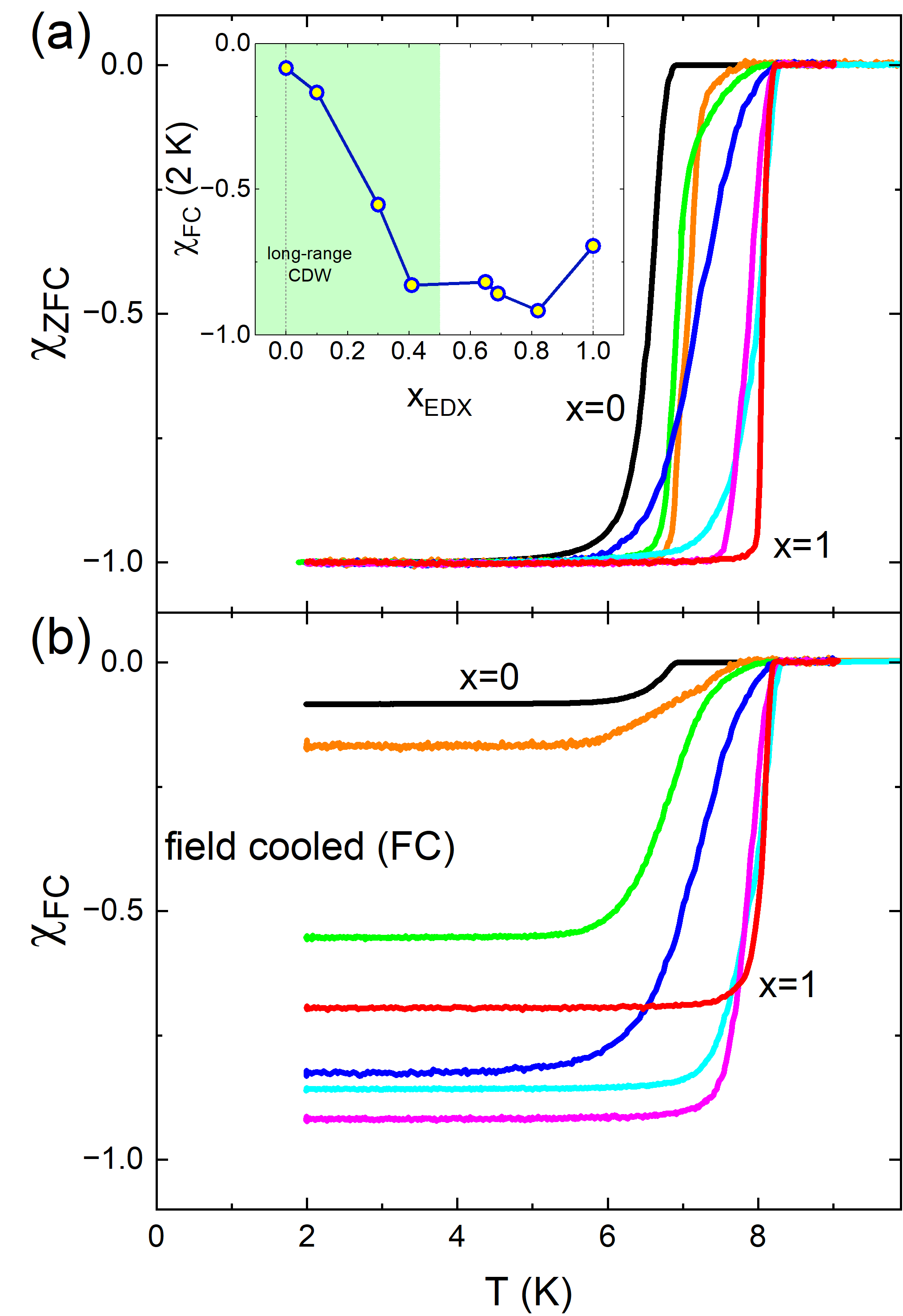} \caption{DC magnetic susceptibility, $\chi=M/H$, of $\text{Ca}_{3}(\text{Ir}_{1-x}\text{Rh}_{x})_{4}\text{Sn}_{13}$
crystals. (a) After cooling in zero magnetic field, applying a 10
Oe field and taking the measurements on warming (ZFC); (b) taking
measurements on cooling in a magnetic field of 10 Oe. }
\label{fig3:mag} 
\end{figure}

DC magnetic susceptibility was measured using a \emph{Quantum Design}
vibrating sample magnetometer (VSM) installed in a 9 T PPMS. Figure~\ref{fig3:mag}(a)
shows magnetic susceptibility, $\chi=M/H$, measured on warming after
the sample was cooled in zero magnetic field and a magnetic field
of 10 Oe was applied, following so-called zero-field-cooled (ZFC) measurement protocol.
In order to compare between the samples that have different volumes
and, importantly, shapes (hence demagnetizing factors), the ZFC data
were normalized to give $\chi=-1$ at the lowest temperature, 2 K
here. With this normalization, the data obtained on cooling in the
same magnetic field of 10 Oe (field-cooling (FC) measurement protocol),
represent the amount of flux expelled by the Meissner effect. This
in turn reflects the amount of pinning in these generally low-pinning
compounds where reversible magnetization dominates the irreversible
one. A detailed study of DC magnetization and radio-frequency Campbell
penetration depth will be published elsewhere \cite{MagCampbell2023}.
Here we focus on the trends observed in the FC state. The values of
magnetic susceptibility at $T=2$ K after FC process are plotted as
function of the Rh composition, $x$, in the inset in Fig.\ref{fig3:mag}(a). Remarkably,
there is a clear correlation with the region of charge-density wave.
More specifically, the expulsion is progressively smaller when CDW
is stronger, judging from the CDW transition temperature (see Fig.\ref{fig6:phase-dia} below).
This points to stronger pinning in the CDW domain, which implies a
direct link between CDW (or structural distortion) and the superconducting condensation energy. Therefore, we can expect a significant effect
if the CDW transition line $T_{CDW}\left(x\right)$ terminates at
$T=0$ as a second order quantum phase transition resulting in a quantum
critical point, which was suggested in other related systems, notably
$(\text{Sr}_{1-x}\text{Ca}_{x})_{3}\text{Ir}_{4}\text{Sn}_{13}$ \cite{Klintberg2012,Goh2015,Goh2018} 
and 
$(\text{Sr}_{1-x}\text{Ca}_x)_3\text{Rh}_4\text{Sn}_{13}$
\cite{Goh2015,Goh2018}.

\subsection{London penetration depth}

\begin{figure}[tb]
\includegraphics[width=7cm]{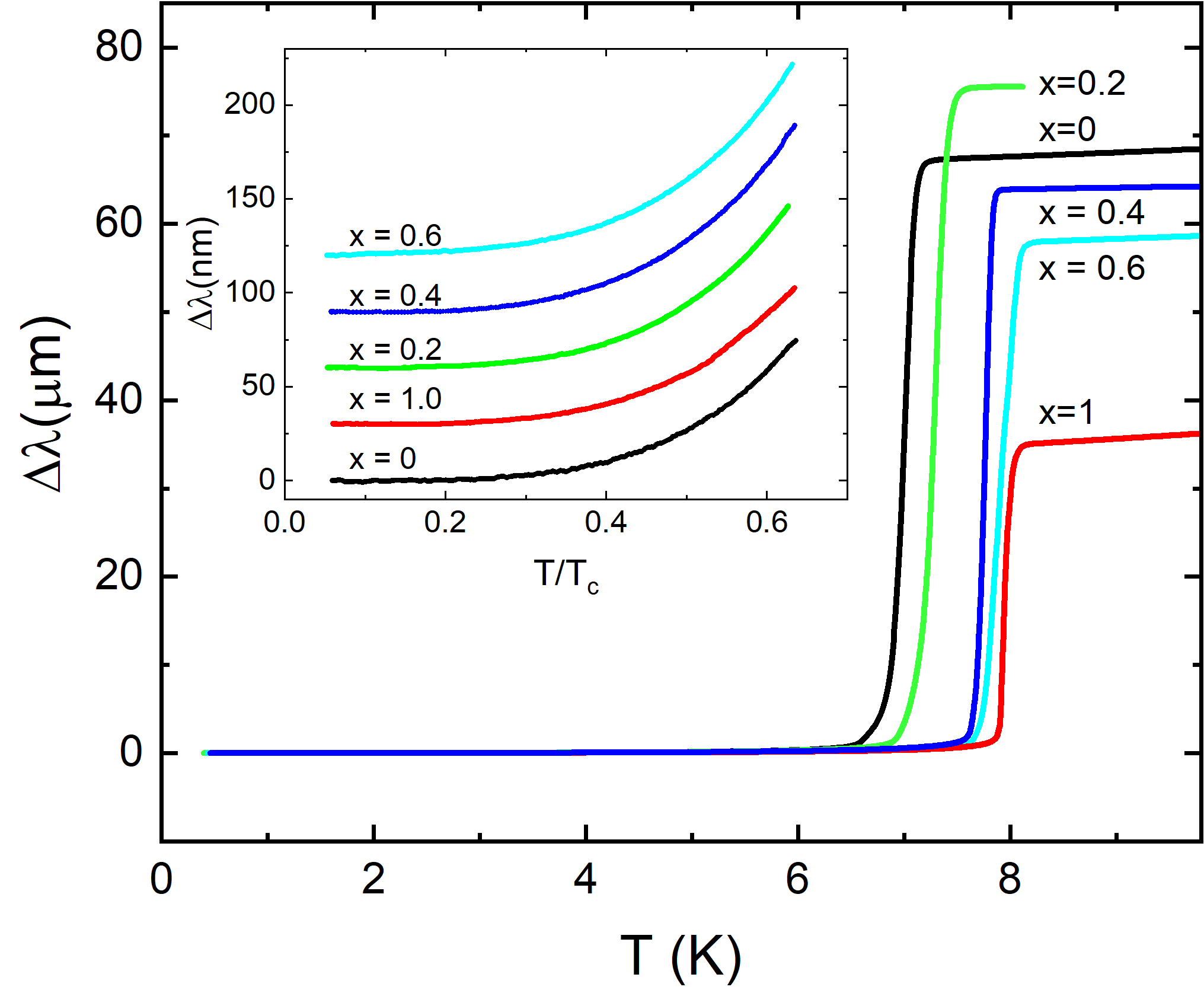} \caption{London penetration depth in $\text{Ca}_3(\text{Ir}_{1-x}\text{Rh}_x)_4\text{Sn}_{13}$
single crystals of indicated Rh compositions. The main panel shows
full temperature range transitions. The saturation just above the
transition accurs when penetration depth starts to diverge and becomes
of the order of the sample size. Due to different sample sizes, there
is no systematic behavior here. The inset zooms at the low-temperature
region, this time plotted as function of the reduced temperature,
$T/T_{c}$. The curves are shifted vertically by a constant for clarity.
Roughly below $0.35T/T_{c}$ all curves saturate indicating exponential
attenuattion, hence a fully gapped state, at all $x$. }
\label{fig4:LPD} 
\end{figure}

The London penetration depth provides further insight into the magnetic
properties of $\text{Ca}_3(\text{Ir}_{1-x}\text{Rh}_x)_4\text{Sn}_{13}$.
The stoichiometric compounds were systematically studied previously \cite{Krenkel2022}. A full description of this unique technique and
its application can be found elsewhere \cite{VanDegrift1975RSI,Prozorov2000PRB,Prozorov2000a,Prozorov2006,Prozorov2021}. In brief, an $LC-$tank
circuit is connected in series with a tunnel diode which when properly
biased has a negative differential resistance. As a result, an optimized
circuit starts resonating spontaneously upon cooling and is always
``locked'' onto its resonant frequency. A superconducting sample
on a sapphire rod is inserted into the inductor without touching it,
in vacuum, so its temperature can be changed without disturbing the
resonator. Mutual magnetic inductive coupling causes a change of the
total magnetic inductance of the circuit, hence the resonant frequency
shift, which is the measured quantity. It can be shown that for each
sample, frequency shift, $\Delta f=-G\chi$, where $\Delta f$ is
measured with respect to the value without the sample (empty coil). Details of
the calibration procedure and calibration constant $G$ are described
elsewhere \cite{Prozorov2006}. The magnetic susceptibility in a Meissnerr-London
state (no vortices) of a superconductor of any shape can be described
by, $\chi=\lambda/R\tanh\left(R/\lambda\right)-1$, where $\lambda$
is the London penetration depth and $R$ is the so-called effective
dimension, which is a function of real sample dimensions \cite{Prozorov2006}.
For typical samples of this research, $R\sim100-200\;\mu\mathrm{m}$.
Therefore, for most of the temperature interval, we can set $\tanh R/\lambda\approx1$
and, therefore, $\delta f\left(T\right)\sim\Delta\lambda\left(T\right)$,
where $\delta f$ is counted from the state at the base temperature,
because we are only interested in the low-temperature variation of
$\lambda(T)$. The circuit stability is such that we resolve the changes
in frequency of the order of 0.01 Hz, which taking into account the
main frequency of 14 MHz mean we have a resolution of 1 part per billion.
For our crystals, this translates to sub-angstrom level sensitivity.

Figure \ref{fig4:LPD} shows London penetration depth in $\text{Ca}_3(\text{Ir}_{1-x}\text{Rh}_x)_4\text{Sn}_{13}$
crystals with indicated Rh content. The main panel shows the full-temperature range with sharp transitions to the normal state. The saturation just
above the transitions occurs when the penetration depth starts to diverge
approaching $T_{c}$ and becomes comparable to the sample size. As
described above, in this case $\lim_{x\rightarrow\infty}\left[\chi=x\tanh\left(1/x\right)-1\right]\rightarrow0$
and measurable signal becomes insensitive to further changes. Comparing
with Fig.\ref{fig3:mag}(a) we find similar-looking curves, because
they both depict $\chi\left(T\right)$. However, if we are to zoom
Fig.\ref{fig3:mag}(a) to low temperatures, we would only find noise,
because commercial susceptometers are only sensitive at about 1 part
per million. In case of TDR, with three orders of magnitude better
sensitivity, we can study the structure of $\lambda(T),$ which is
linked directly to the superconducting order parameter \cite{Prozorov2006}.
The inset in Figure$\;$\ref{fig4:LPD} zooms in on the low-temperature
region, this time plotted as function of the reduced temperature,
$T/T_{c}$. The curves are shifted vertically by a constant for clarity.
Roughly below $0.35T/T_{c}$ all curves saturate indicating exponential
attenuation, hence a fully gapped state, at all $x$. Previously,
we reported exponential attenuation in two limiting pure compounds,
$x=0$ and $x=1$ \cite{Krenkel2022}. Probing the superconducting state robustness
to disorder, we showed a significant reduction of the transition temperature
with non-magnetic scattering and it was concluded that 3-4-13 stannides
are indeed fully gapped, but their order parameter is unconventional
\cite{Krenkel2022}. It seems this trend persists uniformly in the Ir/Rh alloys
opening up possibilities to study various effects, such as nonexpoinential
attenuation of the London penetration depth in the vicinity of the
quantum critical point \cite{Levchenko2022QCP}.

\section{Analysis and discussion}

\begin{figure}[tb]
\includegraphics[width=7cm]{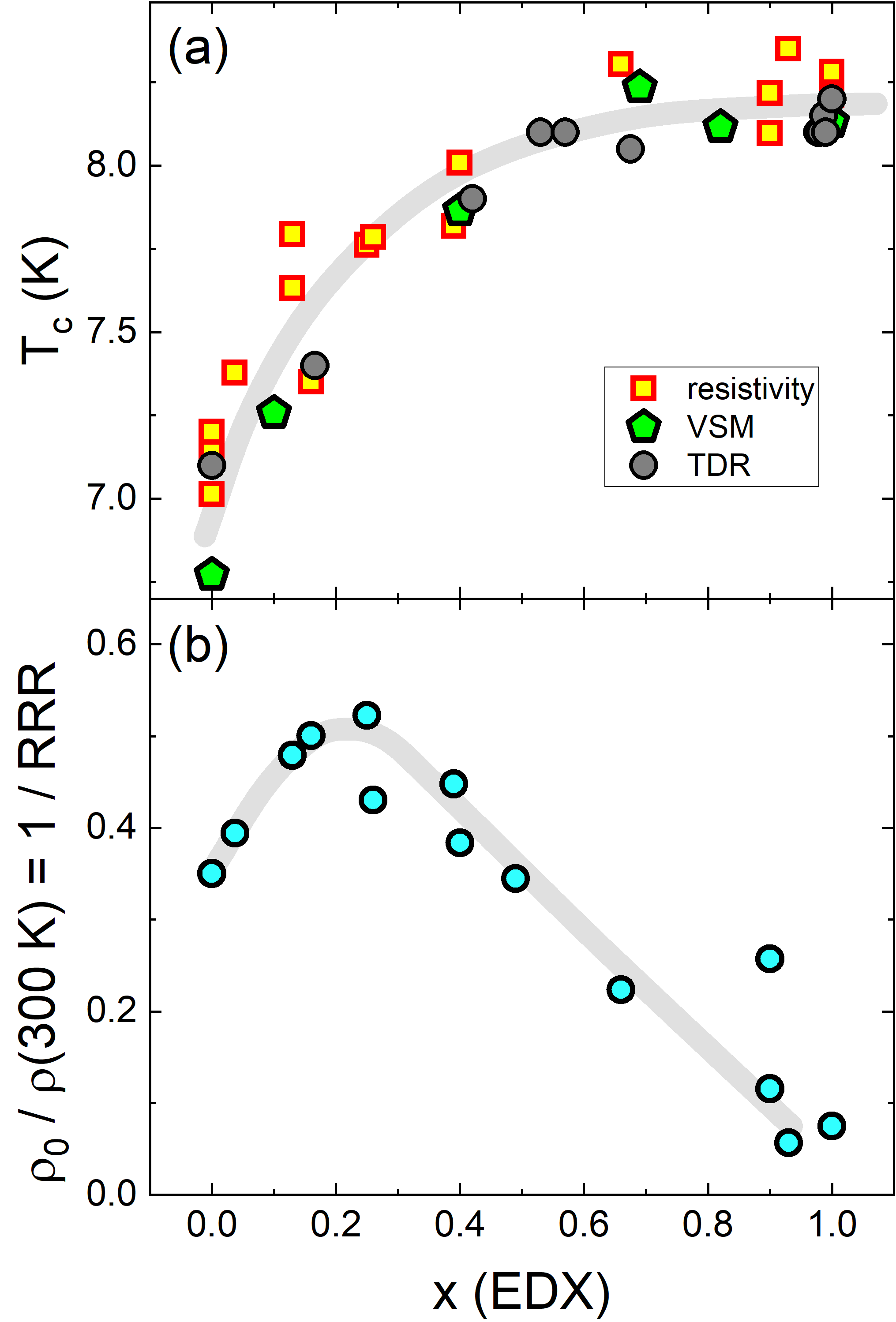} 
\caption{ Composition dependence of the (a) superconducting transition temperature obtained from three independent
measurements, electrical resistivity (red-yellow squares), DC magnetization (black-green pentagons) and tunnel-diode
resonator (dots); and of (b) resistivity above the superconducting $T_c$ normalized by
the resistivity value at the room temperature, $\rho (T_c)/\rho(300K)$.
}
\label{fig5:tc}
\end{figure}

We now summarize and overview the obtained results, focusing on the
trends across the entire range of compositions from pure Ir ($x=0$)
to pure Rh ($x=1$). Figure \ref{fig5:tc}(a) shows the superconducting
transition temperature obtained from three independent measurements,
- electrical resistivity, DC magnetization and tunnel-diode resonator. An increase up to $x \sim$0.5 reflects  a competition
between CDW and superconductivity, though total $T_c$ variation is quite small from from 7 to 8.3~K. A flat plateau in $T_c(x)$ dependence begins above roughly $x=0.5$ without any clear ``dome shape'' peaking of superconductivity around potential quantum critical point, contrary to observations in curpates \cite{Louis}, heavy fermion \cite{Mathur1998} and iron-based superconductors \cite{KasaharaQCP}.

Figure \ref{fig5:tc}(b) shows residual
resistivity immediately above superconducting $T_c$ normalized by the resistivity value at the room temperature, $\rho(T_c)/\rho(300K)$ as a function
of Rh concentration. Expectedly, the resistivity is higher  in the composition range of $x$  where CDW coexists and competes
with superconductivity. This observation clearly shows that two orders are not independent and affect each other. The same conclusion comes from our observation of enhanced pinning discussed above. Of note that residual resistivity continues to decrease with $x$ approaching 1, perhaps reflecting the natural tendency of the system to form Rh-rich phases.

\begin{figure}[tb]
\includegraphics[width=7cm]{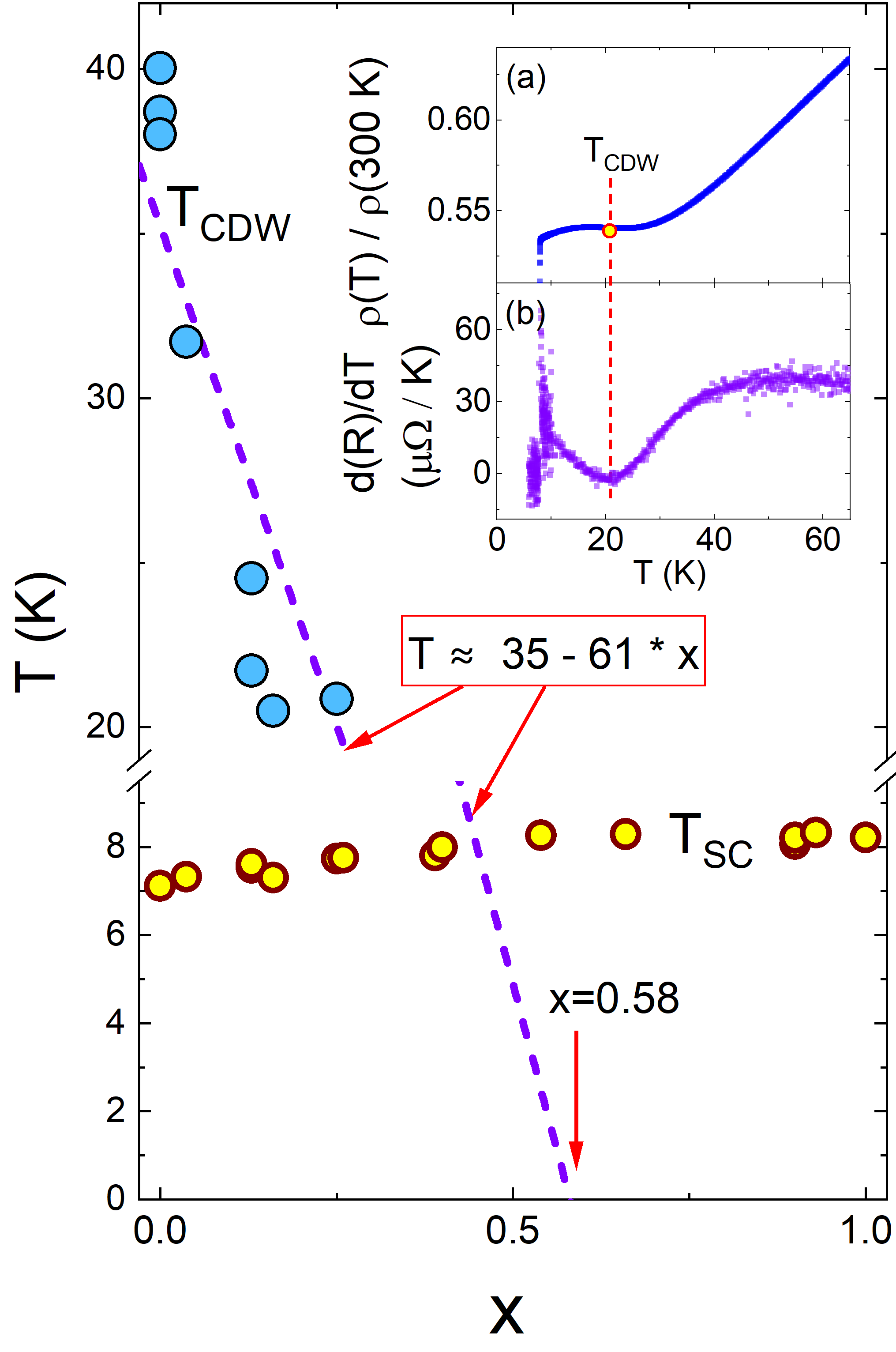} \caption{Summary phase diagram of $\text{Ca}_3(\text{Ir}_{1-x}\text{Rh}_x)_4\text{Sn}_{13}$
alloy.  Blue dots show the CDW transition temperature, $T_{CDW}(x)$,
determined from the derivative of resistivity as shown in the top insets (a)  and (b). The dashed line in the main panel approximates linearly, $T_{CDW}(x)\approx35-61x$, 
to $x=0.58$ revealing the location of a putative quantum critical
point. Note break in the temperature axis.
Inset (a) shows an example of the normalized resistivity vs. temperature
(at $x=0.22$) exhibiting a pronounced upturn upon cooling due to opening of the CDW gap on part of the Fermi surface. The inset shows temperature dependence of the resistivity derivative $d(\rho/\rho(300K)/dT$, with dashed line in the insets showing position of $T_{CDW}$.  Yellow circles in the main panel show composition evolution of the superconducting transition temperature.}
\label{fig6:phase-dia}
\end{figure}

Finally, we present the summary phase diagram of $\text{Ca}_3(\text{Ir}_{1-x}\text{Rh}_x)_4\text{Sn}_{13}$
alloy. Blue dots in Fig.\ref{fig6:phase-dia} show the charge-density-wave
transition temperature, $T_{\text{CDW}}(x_{\text{EDX}})$, determined from the derivative
of resistivity as shown in the inset. Inset (a) shows one of the normalized
resistivity vs. temperature curves ($x=0.22$) with a clear upturn
upon cooling, expected due to opening of the partial gap on the Fermi surface by the CDW. Inset
(b) shows temperature derivative of this curve. The location of $T_{\text{CDW}}(x)$
is taken from the minimum of the derivative. The dashed line approximates the CDW transition data as a line, $T_{\text{CDW}}(x)\approx35-61x$; it extrapolates to zero at $x=0.58$
revealing the possible location of a putative quantum critical point.  

The usual expectation for the behavior of the temperature-dependent resistivity in the vicinity of quantum critical point is to have strong deviations from the $T^2$ dependence expected for a Landau Fermi liquid \cite{Mathur1998,Louis}. This is indeed found in 3-4-13 compositions tuned by pressure \cite{Klintberg2012} and Ca-Sr alloying and pressure \cite{Goh2015}.  Inspection of the resistivity curves in Fig.~\ref{fig2:res}(c) and (d) indeed find $\rho(T)$ which is close to $T$-linear, particularly in the sample $x=$0.9. However, there are still visible deviations with a tendency for upward curvature at the temperatures right above the superconducting transition. It is plausible that the composition variation in the samples makes a mixture of $T$-linear and $T^2$ dependences and masks a true QCP. 

Another expectation is centering of the superconducting $T_c$ with maximum at QCP.  Yellow circles in Fig.\ref{fig6:phase-dia} show superconducting transition
temperature. On this scale, it appears practically independent of
$x$. However, as is shown in Fig.\ref{fig5:tc} it flattens for $x\geq$0.6. 

Finally, the temperature dependent resistivity for all compositions without clear anomalies due to long range CDW ordering in Fig.~\ref{fig2:res}, (e) and (f), show a tendency for saturation at high temperatures. A very similar type of saturation is found in 2H-TaS$_2$, 2H-TaSe$_2$ and their alloys \cite{Naito1,Naito2,Cedomir}. It is possible that charge density wave ordering remains intact, but becomes short-range even in Ca$_3$Rh$_4$Sn$_{13}$. This type of evolution is found in NbSe$_2$ after electron irradiation \cite{KyuilNbSe2}.

\section{Conclusions}

In conclusion, high-quality single crystals of $\text{Ca}_3(\text{Ir}_{1-x}\text{Rh}_x)_4\text{Sn}_{13}$ show a superconducting transition
temperature that increases from 7 K ($x=0$) to 8.3 K ($x=1$). The charge-density wave (CDW) transition is suppressed with composition, $x$, extrapolating linearly to $x_c=0.58$ under the dome of superconductivity. Magnetization and transport measurements show a significant influence of CDW on the superconducting phase. In particular, vortex pinning strength is enhanced in the CDW region, and the normal state resistivity is larger in this part of the
phase diagram. The superconducting $T_c$ does not peak around $x_c$ but rather saturates at $x>x_c$. London penetration depth is attenuated exponentially upon cooling for all compositions, indicating a fully-gapped superconducting state. Overall,
$\text{Ca}_3(\text{Ir}_{1-x}\text{Rh}_x)_4\text{Sn}_{13}$
appears to be a suitable system for finding a quantum critical point at ambient pressure. Supporting this idea is the observation of a $T-$linear temperature-dependent resistivity for $x>x_c$.

\begin{acknowledgments}
This work was supported by the National Science Foundation under Grant No. DMR-2219901.
C.P. acknowledges support by the U.S. Department of Energy, Basic Energy Sciences, Division of Materials Science and Engineering, under Contract No. DE-SC0012704 (BNL). 
\end{acknowledgments}

%\bibliographystyle{apsrev4-2}
%\bibliography{stannides.bib}

\end{document}